\documentclass[aps,prb,groupedaddress,showpacs,twocolumn,letterpaper]{revtex4}
\usepackage{graphics}
\usepackage{epsfig}
\usepackage{graphicx}
\usepackage{dcolumn}

\begin{document}

\title{Spin polarization control by electric field gradients}
\author{Dan Csontos\footnote{Now at Massey University, Palmerston North, New Zealand. E-mail: d.csontos@massey.ac.nz}}
\affiliation{Department of Physics and Astronomy, and Nanoscale and
  Quantum Phenomena
Institute, Ohio University, Athens, Ohio 45701-2979}
\author{Sergio E. Ulloa}
\affiliation{Department of Physics and Astronomy, and Nanoscale and
  Quantum Phenomena
Institute, Ohio University, Athens, Ohio 45701-2979}

\begin{abstract}
{\footnotesize We show that the propagation of spin polarized
carriers may be dramatically affected by {\em inhomogeneous}
electric fields. Surprisingly, the spin diffusion length is found to
strongly depend on the sign and magnitude of electric field
\emph{gradients}, in addition to the previously reported dependence
on the sign and magnitude of the electric field [Yu \& Flatt\'e,
Phys. Rev. B {\bf 66}, 201202(R) 2002; {\em ibid.} {\bf 66}, 235302
(2002)]. This indicates that purely electrical effects may play a
crucial role in spin polarized injection, transport and detection in
semiconductor spintronics. A generalized drift-diffusion equation that describes our findings is derived and verified by numerical calculations using the Boltzmann transport equation.}
\end{abstract}

 \maketitle

\section{Introduction}

Semiconductor spintronics has recently received a lot of attention following the successful demonstration of spin injection from a diluted magnetic semiconductor (DMS)\cite{DMS} to a nonmagnetic semiconductor (NMS).\cite{fiederlingohno} This enables the realization of all-semiconductor spintronics, which potentially can lead to the development of multi-functional devices based on optical and transport processes that make use of both the charge and spin degrees of freedom. However, in order to realize spintronic devices, several prerequisites need to be generally satisfied and achieved: i) efficient creation and injection of spin polarized carriers, ii) transport of the spin-polarized electrons from one location in the device to another, iii) successful manipulation of the spin current signal, iv) efficient detection of the spin-polarized electrons.

While the electron spin is of central importance in the
understanding and design of semiconductor spintronics, the charge of
the electron and its influence on spin transport cannot be
neglected. Charge carrier interactions, drift and diffusion are
expected in any realistic system due to externally applied, or
intrinsic, built-in electric fields, doping concentration variations and
material design. It has been previously shown by Yu and Flatt\'e
(YF) that the spin diffusion length depends on the
magnitude and sign of the electric field.\cite{yu} Such a dependence was
recently experimentally observed.\cite{crookerPRL2005} It has also
been shown experimentally that spin-polarized transport and
injection is very sensitive to an applied bias
voltage,\cite{schmidtPRL2004, vandorpeAPL2004, adelmannJVAC2005,
kohdaCONDMAT2005,crookerPRL2005} as well as different structural
parameters such as doping concentrations\cite{adelmannJVAC2005} and
layer thicknesses.\cite{kohdaCONDMAT2005} These considerations pose
the fundamental question: What is the influence of purely electrical
effects on spin injection, transport and detection in semiconductor
spintronics?

Here we show that the the spin diffusion length may have a dramatic
dependence on electric field {\em gradients}, an effect that can be
significantly stronger than the electric field effects predicted by
YF.\cite{yu} In essence, our theoretical and numerical study
predicts that even small deviations from charge neutrality can give
rise to significant enhancement or suppression of spin polarization.

Previously, it has been assumed that the effects of inhomogeneous
fields, such as, e.g., occurring at material or doping concentration
interfaces, should be relatively small, since the charge screening length
typically is much shorter than the spin diffusion length. We show
that the effective spin diffusion length in fact can be comparable
to the screening length in the presence of inhomogeneous electric
fields. In particular, we study the propagation of spin polarized
electrons across interfaces between regions with different doping
concentrations. The spin polarization of electrons moving from a
region with lower to a region with higher doping concentration is
found to be strongly suppressed due to the electric field gradient
effect. Conversely, electric field gradients may enhance the spin
diffusion length, depending on their sign and magnitude. 

Our results
show that purely electrical effects may have a profound impact on
spin injection, spin transport, and spin detection, three key
ingredients for the realization of semiconductor spintronics. A
generalized drift-diffusion equation which is able to describe
nonequilibrium spin density propagation in the presence of
inhomogeneous electric fields is derived and verified with numerical
calculations using the Boltzmann transport equation (BTE). We note that recent theoretical studies by, e.g., Wang and Wu\cite{wu} and Sherman,\cite{sherman} have also predicted very interesting effects related to electric-field induced spin relaxation mechanisms due to the Rashba effect. 

In the following, we will first describe our self-consistent numerical approach (Section \ref{numerics}), followed by numerical results from which the strong influence of electric fields on spin transport will become evident (Section \ref{results}). Subsequently, in Section \ref{DDequations} we derive spin drift-diffusion equations which are then used to discuss and understand the numerically observed characteristics (Section \ref{discussion}). Finally, our conclusions are presented in Section \ref{conclusions}. 

\section{Self-consistent Boltzmann-Poisson model}
\label{numerics}
In order to fully understand spin transport properties in
semiconductor structures, and in particular the influence of
interfaces and inhomogeneous electric fields, we take into account
nonequilibrium transport processes for {\em both} the {\em charge}
and {\em spin} degrees of freedom in a self-consistent way. In our
approach, the transport of spin-polarized electrons is described by
two BTE equations according to
\begin{equation}
\label{BTE} -\frac{e}{m^{\ast}}{\mathbf E}\cdot
\mathbf{\nabla}_{\mathbf v}f_{\uparrow (\downarrow)} + {\mathbf v}
\cdot \mathbf{\nabla}_{\mathbf{r}}f_{\uparrow(\downarrow)} =
-\frac{f_{\uparrow(\downarrow)}-f^{0}_{\uparrow(\downarrow)}}{
\tau_{{\mathrm m}}} - \frac{f_{\uparrow (\downarrow)}-f_{\downarrow
(\uparrow)}} {\tau_{\uparrow \downarrow (\downarrow \uparrow)}}~,
\end{equation}
where $\mathbf{E}$ is an inhomogeneous electric field,
$f_{\uparrow(\downarrow)}$ is the electron distribution for the
spin-up (down) electrons,  $\tau_{{\mathrm m}}$ is the momentum
relaxation time, and $1/\tau_{\uparrow \downarrow}$
($1/\tau_{\downarrow \uparrow}$) is the rate at which spin-up
(spin-down) electrons scatter to spin-down (spin-up) electrons. The
first term on the right-hand side of eq.\ (\ref{BTE}) describes the
relaxation of each nonequilibrium spin distribution to a local
equilibrium (spin-dependent) electron distribution function,
$f^{0}_{\uparrow(\downarrow)}$, which we choose as non-degenerate according to 
\begin{equation}
f^{0}_{\uparrow (\downarrow)}= n_{\uparrow
(\downarrow)}(\mathbf{r}) \sqrt{\frac{m^{\ast}}{2\pi k_{B}T}}
\exp{(-m\mathbf{v}^{2}/2k_{B}T)}~,
\label{leq}
\end{equation} 
where $T$ is the lattice
temperature. The last term in eq.\ (\ref{BTE}) describes the
relaxation of the spin polarization. From the distribution function
$f_{\uparrow (\downarrow)}$ we calculate the local spin density
according to 
\begin{equation}
n_{\uparrow (\downarrow)}(\mathbf{r})= \int
f_{\uparrow (\downarrow)}(\mathbf{r},\mathbf{v})d{\mathbf v}~,
\label{spindensity}
\end{equation}
and
define two quantities, the spin density {\em imbalance}
\begin{equation}
\delta_{\uparrow \downarrow}=n_{\uparrow}-n_{\downarrow}~,
\end{equation} 
where
$n_{\uparrow}$ and $n_{\downarrow}$ are the densities of spin-up and
spin-down polarized electrons, and the spin density
\emph{polarization}
\begin{equation}
P_{n}=\frac{n_{\uparrow}-n_{\downarrow}}{n_{\uparrow}+n_{\downarrow}}=
\frac{\delta_{\uparrow \downarrow}}{n}~,
\end{equation} 
where
$n=n_{\uparrow}+n_{\downarrow}$ is the total charge density.

Inhomogeneous charge distributions and electric fields are taken
into account by solving the Poisson equation
\begin{equation}
{\mathbf \nabla}^{2}\phi=-{\mathbf \nabla} \cdot {\mathbf E} = -e\frac{N_{D} - n_{\uparrow} -
n_{\downarrow} }{\varepsilon \varepsilon_{0}}~, \label{poisson}
\end{equation}
where $\varepsilon$ is the dielectric constant, $N_{D}$ is the
donor concentration profile, and $\phi$ and $E$ are the spatially dependent potential and electric field profiles. 

Our
theoretical model thus goes beyond drift-diffusion, and is capable
of describing charge and spin transport through strongly
inhomogeneous (with respect to spin and charge densities)
semiconductor systems, as well as nonequilibrium effects. We assume
in the following that an electric field is applied along the $x$
direction and consider the corresponding one-dimensional transport
problem. 
\begin{center}
\begin{figure}
\scalebox{1}{\epsfig{file=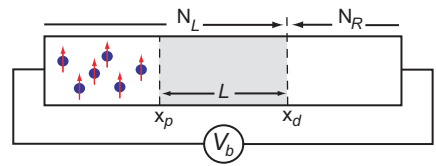}}
\caption{\label{figure1} (color
online) Schematics of the model structure we consider. Full spin
polarization, $P_{n}=1$ is assumed for $x<x_{p}$.}
\end{figure}
\end{center}
The nonlinear, coupled equations
(\ref{BTE},\ref{leq},\ref{spindensity},\ref{poisson}) need to be solved self-consistently for a
given applied bias voltage, doping concentration profile,
nonequilibrium spin distributions, and relaxation times. We use a numerical approach based on finite difference and relaxation methods, that we originally developed for the study of nonequilibrium effects in charge transport through ultrasmall, inhomogeneous semiconductor channels.\cite{csontosJCE, csontosAPL} As boundary conditions, we adopt the following scheme: For the potential, the values at the system boundaries are fixed to $\phi(x_{l})=V_{b}$ and $\phi(x_{r})=0$ ($l,r$ denote the left and right boundary of the sample, respectively), corresponding to an externally applied voltage $V_{b}$. The {\it electron charge density} is allowed to fluctuate in the system subject to the condition of {\it global} charge neutrality, which is enforced between each successive iteration in the self-consistent Poisson-Boltzmann loop. The {\it spin density} at the DMS boundary and in the DMS is determined by the degree of polarization $P_{n}=(n_{\uparrow}-n_{\downarrow}) /(n_{\uparrow}+n_{\downarrow})$, for which the density is defined according to $n_{\uparrow(\downarrow)}=n/2(1\pm P)$. In the calculations, the size of the contacts has to be large enough, such that the electric field deep inside the contacts is constant and low. This allows us to use the analytical, linear response solution to the BTEs (\ref{BTE})
\begin{equation}
f_{\uparrow(\downarrow)}(x_{l,r},v)=f^{0}_{\uparrow(\downarrow)}(x_{l,r},v) [1-vE(x_{l,r})\tau_{m}/k_{B}T]~,
\end{equation}
as boundary conditions at $x_{l,r}$, where we use the local equilibrium distribution $f^{0}_{\uparrow(\downarrow)}(x_{l,r},v)$ and local electric field, $E(x_{l,r})$, obtained from the previous numerical solution to the Poisson-Boltzmann iterative loop. More details of the numerical procedure for a similar problem are given in Ref.\ \onlinecite{csontosPHYSICAE}.

The model structure that we study is described in Fig.\ \ref{figure1}. As
seen in the figure, the system is inhomogeneous with respect to both
charge and spin degrees of freedom. Spin-polarized electrons, with
$P_{n}=1$, are injected from a diluted magnetic
semiconductor(DMS)-like portion,\cite{DMS, fiederlingohno} defined
for $x<x_{p}$, into a nonmagnetic semiconductor (NMS) part. The
spin-polarized electrons subsequently relax due to spin-flip
scattering for $x>x_{p}$, at the rate $1/\tau_{{\mathrm sf}}$ (where
$\tau_{\uparrow \downarrow}=\tau_{\downarrow
\uparrow}=2\tau_{{\mathrm sf}}$), to the asymptotic unpolarized
value $P_{n}=0$ for $x\gg x_{p}$. The doping concentration is
defined as $N_{D}=N_{L}$ and $N_{D}=N_{R}$ for $x<x_{d}$ and $x\geq
x_{d}$, respectively. The following material and system parameters have been used in our
study: $T=300$ K, $m^{\ast}=0.067$m$_{0}$, $\tau_{{\mathrm m}}=0.1$
ps (typical for GaAs at room temperature), $\tau_{{\mathrm sf}}=1$
ns, $N_{L}=10^{21}$ m$^{-3}$, $V_{b}=-0.3$ V. The total system size is 5 $\mu$m, i.e., $x_{l,r}=\mp 2.5$ $\mu$m.

\section{Numerical results: Enhancement and suppression of spin polarization at doping interfaces}
\label{results}
\begin{center}
\begin{figure}
\scalebox{1}{\epsfig{file=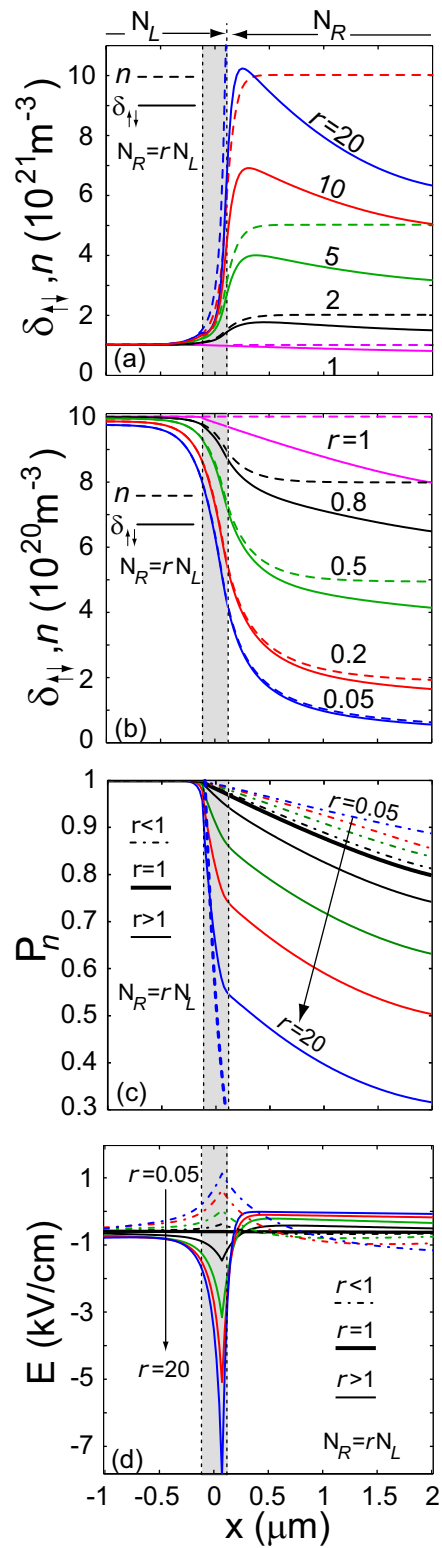}}
\caption{\label{figure2}(color online) Spin density imbalance
$\delta_{\uparrow \downarrow}$ (solid lines) and total charge density $n$ (dashed lines) for a spin and charge inhomogeneous
structure as depicted in Fig.\ \ref{figure1} for different doping concentrations (a) $N_{R}=1,2,5,10,20\cdot N_{L}$, and (b) $N_{R}=0.05,0.2,0.5,0.8,1\cdot N_{L}$ (increasing from bottom to top). (c) Spin density polarization $P_{n}$, and (d) electric field distribution for the same structure. The structure
parameters are $N_{L}=10^{21}$ m$^{-3}$, $L=0.2$ $\mu$m,
$\tau_{{\mathrm sf}}=0.5$ ns, $\tau_{{\mathrm m}}=0.1$ ps,
$V_{b}=-0.3$ V.}
\end{figure}
\end{center}

In the following, we present numerical results for a spin and charge inhomogeneous semiconductor structure and highlight the characteristics of electric field effects on the spin polarization. In Fig.\ \ref{figure2} we study the propagation of spin-polarized
electrons across two spin- and charge-inhomogeneous interfaces,
separated by $L=0.2$ $\mu$m (grey mid-region), as a function of
different doping concentrations $N_{R}$. In Fig.\ \ref{figure2}(a)
we show the spin density imbalance, $\delta_{\uparrow
\downarrow}=n_{\uparrow}-n_{\downarrow}$ (solid lines), and the
total charge density, $n$ (dashed lines), for $N_{R}=rN_{L}$, where
$r=1,2,5,10,20$ ($N_{L}=10^{21}$ m$^{-3}$. The lowest solid and dashed curves correspond to $\delta_{\uparrow\downarrow}$ and $n$ of a
charge homogeneous sample, $N_{R}=N_{L}$, in which the electric field is
accordingly constant, as shown by the thick, solid line in Fig.\
\ref{figure2}(d). In contrast to the constant total charge density (bottom dashed line), for the charge homogeneous case, the spin density imbalance,
$\delta_{\uparrow \downarrow}$ (bottom solid line), decreases monotonically for
$x>x_{p}$, which coincides with the position of the DMS/NMS interface and the onset of the spin flip scattering rate $1/\tau_{{\mathrm sf}}$. The decay is exponential
and can be understood in terms of the drift-diffusion description in
the YF model.\cite{yu} Within this model, the spin density imbalance decay can be described by $\delta_{\uparrow \downarrow} \sim
\exp[-x/L_{D(U)}]$ where
\begin{equation}
L_{D(U)}=\left\{ -(+)\frac{\left\vert eE\right\vert
}{2k_{B}T}+\sqrt{\left( \frac{eE}{2k_{B}T}\right)
^{2}+\frac{1}{L_{s}^{2}}}\right\} ^{-1}~, \label{spindiff}
\end{equation}
and where, $L_{d(u)}$ are electric-field dependent spin diffusion
lengths, describing spin propagation antiparallel (parallel) to the
electric field. In eq.\ (\ref{spindiff}) $L_{s}=\sqrt{D\tau _{sf}}$
is an intrinsic spin-diffusion length in the absence of an electric
field and $D=k_{B}T\tau _{m}/m^{\ast }$, as obtained from the
Einstein relation. From eq.\ (\ref{spindiff}) it follows that the
spin-diffusion length is \emph{enhanced} in the direction
anti-parallel to an applied electric field and \emph{suppressed} in
the direction parallel to the field.\cite{yu} Our numerical results
agree with these considerations for the homogeneous system.

For increasing $N_{R}$, the total charge density, $n$, naturally increases monotonically for $x>x_{d}$, reaching $n\approx N_{R}$ on the order of a screening length to the right of the doping interface. The spin density imbalance $\delta_{\uparrow \downarrow}$, however, has a more complicated behavior. At
the interface between the two doping regions, defined at $x_{d}$, a sharp increase of
$\delta_{\uparrow \downarrow}$ is observed, which subsequently peaks below the maximum value of $n$, followed by a monotonic decay. Similar spin accumulation was recently reported by Pershin and Privman.\cite{pershinPRL2003} We note and emphasize, however, that the {\em spin accumulation peak is significantly lower} than the maximum value of $n$ around the doping interface, e.g., for the $N_{R}=20\cdot 10^{21}$ m$^{-3}$ sample, $\delta_{\uparrow\downarrow}$ reaches only $\approx 50\%$ of the corresponding value in $n$. 

For $N_{R}<N_{L}$, when electrons are injected from a high to a low
doping region, the situation is reversed. In Fig.\ \ref{figure2}(b),
the dashed and solid lines correspond to $n$ and $\delta_{\uparrow
\downarrow}$, calculated for different values of $N_{R}=rN_{L}$,
where $1\geq r\geq 0.05$. For decreasing $N_{R}$, $\delta_{\uparrow
\downarrow}$ (and naturally $n$) decreases as well.

The observed increase (decrease) of $\delta_{\uparrow \downarrow}$
with increasing (decreasing) $N_{R}$ can at first order be understood in terms of the large difference between the intrinsic spin relaxation length, $L_{s}$,
and the Debye screening length, $L_{DB}$. The screening length is
generally shorter than the spin relaxation length. For example, for
the range of doping concentrations considered here, we find
$0.03<L_{DB}<0.6$ $\mu$m, whereas $L_{s}\approx 1.8$ $\mu$m. The
spin density imbalance can be written in terms of $\delta_{\uparrow
\downarrow}=n-2n_{\downarrow}$. Since $L_{s}> L_{DB}$, the total
charge density, $n$, will increase (decrease) faster than
$n_{\downarrow}$ with increasing (decreasing) $N_{R}$. Hence, one would expect the spatial dependence of $n$ to dominate $\delta_{\uparrow
\downarrow}$ within a screening length of the doping interface. This
is clearly seen by comparison between $\delta _{\uparrow
\downarrow}$ (solid lines) and $n$ (dashed lines) in Figs.
\ref{figure2}(a,b).

For $x\gg x_{d}$, $\delta_{\uparrow \downarrow}$ is seen to decrease exponentially. Furthermore, the decay rate is larger for increasing doping concentrations $N_{R}$. One can understand this spatial dependence from previous arguments and the model of YF. For $x\gg x_{d}$ the
total charge density $n$ and the electric field is constant (and thus we can apply the YF model) and the spin-flip scattering drives the nonequilibrium spin density distribution exponentially toward equilibrium,
$\delta_{\uparrow \downarrow}=0$ on a length scale given by eq.
(\ref{spindiff}). Since an increase in $N_{R}$ gives rise to a decrease of the electric field in that region [see Fig.\ \ref{figure2}(d) for the electric profiles] the spin diffusion length correspondingly decreases consistent with the predictions of eq.\ (\ref{spindiff}), giving rise to a faster decay of the spin density imbalance.

The description provided by eq.\ (\ref{spindiff}), however, fails to
describe the magnitude, as well as the spatial dependence of $\delta
_{\uparrow \downarrow}$ around the doping interface. For example,
considering first the curves with $N_{R}>N_{L}$, for $N_{R}=2N_{L}$,
$\delta_{\uparrow \downarrow}$ follows the spatial profile of $n$
rather closely around the interface region. Surprisingly, with
increasing $N_{R}$, the maximum value reached by $\delta_{\uparrow
\downarrow}$ becomes increasingly smaller than $N_{R}$, and the spin density imbalance thus gets strongly suppressed. This is an
unexpected finding, and in contradiction the YF model and with eq.\ (\ref{spindiff}), since, according to this equation, a strong built-in negativ field, such as shown in Fig.\ \ref{figure2}(d), around the doping interface should enhance the spin diffusion length significantly and hence, should yield $\delta_{\uparrow \downarrow}\approx n$ until the electric field drops again for $x>>x_{d}$. 

This discrepancy is also well illustrated in the spin
density {\em polarization}, $P_{n}$, shown in Fig.\
\ref{figure2}(c). Although, as naively expected, $P_{n}=\delta_{\uparrow \downarrow}/n$ drops
sharply around the interface due to the sharp rise in $n$, the
\emph{dependence} on $N_{R}$ is in contradiction with eq.
(\ref{spindiff}) due to the following: An increase in $N_{R}$,
yields a sharp increase of $|E|$ around the interface [see Fig.\
\ref{figure2}(d)], and should therefore lead to a significant
increase of the "downstream" spin diffusion length, $L_{D}$,
according to eq. (\ref{spindiff}). Thus, an increase in $N_{R}$
should be beneficial for $\delta_{\uparrow \downarrow}$ and $P_{n}$,
in contradiction to our findings. For $N_{R}<N_{L}$, a similar
unexpected dependence on $N_{R}$ is observed.

Clearly, one needs to go beyond the YF model (which is valid for a system with local charge neutrality) in order to understand spin polarized transport in the presence of inhomogeneous electric fields . As we will discuss below, the sign and magnitude of the electric field and electric field gradients around the interface are crucial for determining the magnitude and spatial dependence of the spin density imbalance and polarization around the doping interface. 

\section{Spin drift-diffusion equations}
\label{DDequations}

A qualitative understanding of the unexpected spatial dependence of $P_{n}$ and
$\delta _{\uparrow \downarrow}$, can be obtained by reexamining the spin drift-diffusion equations. Previous works have mainly focused on
the charge homogeneous case.\cite{yu, osipovPRB2005, agrawalPRB2005,
deryPRB2006} This is a natural assumption since the screening length
is expected to be much shorter than the intrinsic spin relaxation length,
leading one to assume quasi-local charge neutrality. However, as we
will show below, corrections to the spin drift-diffusion equations of YF\cite{yu} yield that even quasi-local deviations from charge neutrality
can have a very strong impact on the spin polarization.

In the following, we perform the same steps as YF\cite{yu} for the derivation of the spin drift-diffusion equations. The current for spin-up and spin-down electrons can be written as
\begin{equation}
{\mathbf j}_{\uparrow(\downarrow)} =
\sigma_{\uparrow(\downarrow)}{\mathbf E}+ eD_{\uparrow (\downarrow)}
\nabla n_{\uparrow(\downarrow)}~, \label{DD}
\end{equation} where
$D_{\uparrow(\downarrow)}$ and $\sigma_{\uparrow(\downarrow)}$ are
the diffusion constants and conductivities for the spin-up(down)
species. Here, the conductivities refer to
$\sigma_{\uparrow(\downarrow)}=en_{\uparrow(\downarrow)}\mu_{\uparrow(\downarrow)}$, in terms of the mobilities $\mu_{\uparrow(\downarrow)}$ and densities $n_{\uparrow(\downarrow)}$ of spin-up(down) electrons. The continuity equations for the two species require that
\begin{equation}
\frac{\partial n_{\uparrow(\downarrow)}}{\partial t} = -
\frac{n_{\uparrow(\downarrow)}}{\tau_{\uparrow\downarrow(\downarrow\uparrow)}}
+
\frac{n_{\downarrow(\uparrow)}}{\tau_{\downarrow\uparrow(\uparrow\downarrow)}}
+ \frac{1}{e}\nabla \cdot {\mathbf j}_{\uparrow(\downarrow)}~,
\label{continuity}
\end{equation} where
$\tau_{\uparrow\downarrow}(\tau_{\downarrow\uparrow})$ are the
spin-flip scattering rates. In steady-state eqs.
(\ref{DD},\ref{continuity}) result in
\begin{equation}
\nabla \sigma_{\uparrow(\downarrow)}\cdot {\mathbf E} + \sigma_{\uparrow(\downarrow)}\nabla \cdot {\mathbf E} +eD_{\uparrow(\downarrow)}\nabla^{2} n_{\uparrow(\downarrow)} =  -\frac{en_{\uparrow(\downarrow)}}{\tau_{\uparrow\downarrow(\downarrow\uparrow)}} + \frac{en_{\downarrow(\uparrow)}}{\tau_{\downarrow\uparrow(\uparrow\downarrow)}}~. \label{DDc}
\end{equation} 
For a NMS $\mu_{\uparrow}=\mu_{\downarrow}$, and $D_{\uparrow}=D_{\downarrow}$. Multiplying eq.\ (\ref{DDc}) with $\sigma_{\downarrow}(\sigma_{\downarrow})$ 
and substracting them from each other we arrive
to the following expression for the spin density imbalance
$\delta_{\uparrow\downarrow}=n_{\uparrow}-n_{\downarrow}$
\begin{equation}
\nabla^{2} \delta_{\uparrow\downarrow} + \frac{e{\mathbf
E}}{k_{B}T}\nabla \delta_{\uparrow\downarrow} + \frac{e{\mathbf
E}}{2k_{B}T}\delta_{\uparrow\downarrow} \nabla \cdot {\mathbf E} -
\frac{\delta_{\uparrow\downarrow}}{L_{s}}=0~,
\label{DDimbalance}
\end{equation} where we have used the Einstein
relation $\mu=eD/k_{B}T$, $\tau_{s}$ is the spin relaxation time
$\tau_{s}^{-1}=\tau_{\uparrow\downarrow}^{-1}+\tau_{\downarrow\uparrow}^{-1}$
($\tau_{\uparrow\downarrow}=\tau_{\downarrow\uparrow}$), and
$L_{s}=\sqrt{D\tau_{s}}$ is the intrinsic spin diffusion length.
\begin{center}
\begin{figure}
\scalebox{1}{\epsfig{file=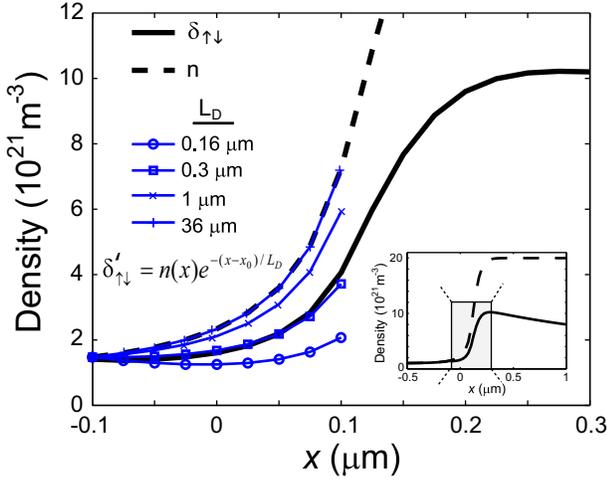}}
\caption{\label{figurePASPS}(color online) Total charge density (dashed line) and spin density imbalance (solid line) profiles calculated for $N_{R}=20N_{L}$. Additional curves for $-0.1<x<0.1$ $\mu$m correspond to $\delta_{\uparrow \downarrow}=n(x)\exp[-(x+0.1)/L_{D}]$ for $L_{D}=0.16,0.3,1,36$ $\mu$m, as described in the text. Inset shows $n$ and $\delta_{\uparrow\downarrow}$ for a larger region.}
\end{figure}
\end{center}
Equation (\ref{DDimbalance}) is a drift-diffusion equation for the
spin density imbalance $\delta_{\uparrow\downarrow}$ and contains
terms depending on both the electric field, and the electric field
{\em gradient}. We note that in fact eq.\ (\ref{DDimbalance})
resembles the corresponding equations in Refs.\ \onlinecite{yu}, but
with an {\em additional term proportional to} $\nabla \cdot {\mathbf
E}$. Similarly to Refs.\ \onlinecite{yu} we can use the roots of
the characteristic equation for eq.\ (\ref{DDimbalance}) to define
up- and down-stream spin diffusion lengths from eq.
(\ref{DDimbalance}) leading up to the following equation
\begin{equation}
L_{D(U)}^{\prime}=\left [-(+)\frac{|eE|}{2k_{B}T} + \sqrt{\left (
\frac{eE}{2k_{B}T}\right )^{2} + \frac{1}{L_{s}^{2}}- \frac{e\nabla
E}{k_{B}T} } \right ]^{-1}~.
\label{LDU}
\end{equation} 

Notice that eq.\ (\ref{LDU}) is only defined "locally", over a region where $\nabla \cdot {\mathbf E}$ can be considered constant, and where an
average value of the electric field is used. Note also that the
above equations are also valid for ferromagnetic semiconductors
provided the mobility and diffusion constant are replaced with the
ones corresponding to the minority-spin species.\cite{yu,
flattePRL}

A comparison between eqs.\ (\ref{LDU}) and (\ref{spindiff}) shows that eq.\ (\ref{LDU}) contains an additional term $-e\nabla E/k_{B}T$ in the square root expression. This
term is a correction to the YF model which describes the important role that inhomogeneous electric fields can play on spin polarized transport. 

\section{Discussion}
\label{discussion}

From eq.\ (\ref{LDU}) it is thus seen that the drift-diffusion length depends strongly on the \emph{gradient}
of the electric field; Consequently, the spin density imbalance and
polarization decay lengths are enhanced or suppressed based not only
on the sign of the electric field alone.\cite{yu} On the contrary,
depending on the profile of the electric field, the gradient term in
eq.\ (\ref{LDU}) can either enhance or suppress the
spin-polarization propagation significantly, as seen explicitly in
our calculated results in Fig.\ \ref{figure2}(a-c).

To illustrate the dramatic impact of the electric field gradient term on the
spin polarization diffusion, we calculate $L^{\prime}_{D(U)}$ for
the central region of the structure, $x_{p}<x<x_{d}$. Using a linear
approximation of the electric field within this region, for the structure with $N_{R}=20\cdot 10^{21}$ m$^{-3}$ [bottom, solid line
in Fig.\ \ref{figure2}(c)], the intrinsic spin-diffusion length
$L_{s}\approx 1.8$ $\mu$m, and an electric field value taken at
mid-channel, $E=-2.9$ kV/cm, eq.\ (\ref{LDU}) yields
$L^{\prime}_{D}\approx 0.16$ $\mu$m and $L^{\prime}_{U}\approx 0.06$
$\mu$m. In comparison, an evaluation of eq.\ (\ref{spindiff}) using
the same value of the electric field, yields $L_{D}\approx 36$
$\mu$m, $L_{U}\approx 0.09$ $\mu$m. Hence, the electric field
gradient in this case decreases the ``local'' (around the interface)
spin diffusion length, along the direction of (charge) transport, by
two orders of magnitude! \emph{The injected spin density
polarization is thus destroyed by the inhomogeneous electric field
at the doping interface.}

To verify the validity of eq.\ (\ref{LDU}), we compare $\exp[-(x-x_{p})/L^{\prime}_{D}]$,
using $L^{\prime}_{D}=0.16$ $\mu$m as before, with the results (for
$P_{n}$) from our Boltzmann-Poisson numerical calculation, within
$x_{p}<x<x_{d}$. The results, plotted in Fig.\ \ref{figure2}(c)
(thick dashed line for the highest $N_{R}$ value), are found
to agree very well with the numerically calculated values. 

We further do a similar analysis for the spin density imbalance, $\delta_{\uparrow\downarrow}$. Here, we compare the numerically calculated values for $\delta_{\uparrow \downarrow}$, and $n$ for the same doping concentration profile with $N_{R}=20N_{L}$, with $\delta^{\prime}_{\uparrow \downarrow} =n(x)\exp[-(x-x_{0})/L^{\prime}_{D}]$, using the numerically calculated values of $n(x)$, and the estimates of $L_{D}$ and $L^{\prime}_{D}$ above. The results are shown in Fig.\ \ref{figurePASPS} for $L^{\prime}_{D}=0.16,0.3,1$ and 36 $\mu$m, the first and latter values corresponding to the previously estimated values using eqs.\ \ref{LDU} and \ref{spindiff}. The numerically calculated curve of $\delta_{\uparrow\downarrow}$ is best fitted to an effective spin diffusion length of $L^{\prime}_{D}=0.3$ $\mu$m, which is in good agreement with the value obtained from eq.\ (\ref{LDU}), any discrepancies being due to the approximations in the choice of $\nabla E$ and $E$ in eq.\ (\ref{LDU}). Nevertheless, it is clear that the spin diffusion length around the interface is much smaller than the intrinsic one, $L_{s}$, and the corresponding $L_{D}$, obtained within the YF model, which is only valid in the (locally) charge neutral case.

Reversing the sign of the electric field gradient, the spin density
polarization $P_{n}$ is in fact \emph{enhanced} in comparison to the
homogeneous case. This is illustrated by the dashed-dotted curves in
Fig.\ \ref{figure2}(c), which correspond to $N_{R}<N_{L}$. Since the
electric field gradients are not as large as for the previously
studied case [see Fig.\ \ref{figure2}(d)], the corresponding
increase in $P_{n}$ is more modest.

\begin{center}
\begin{figure}
\scalebox{1}{\epsfig{file=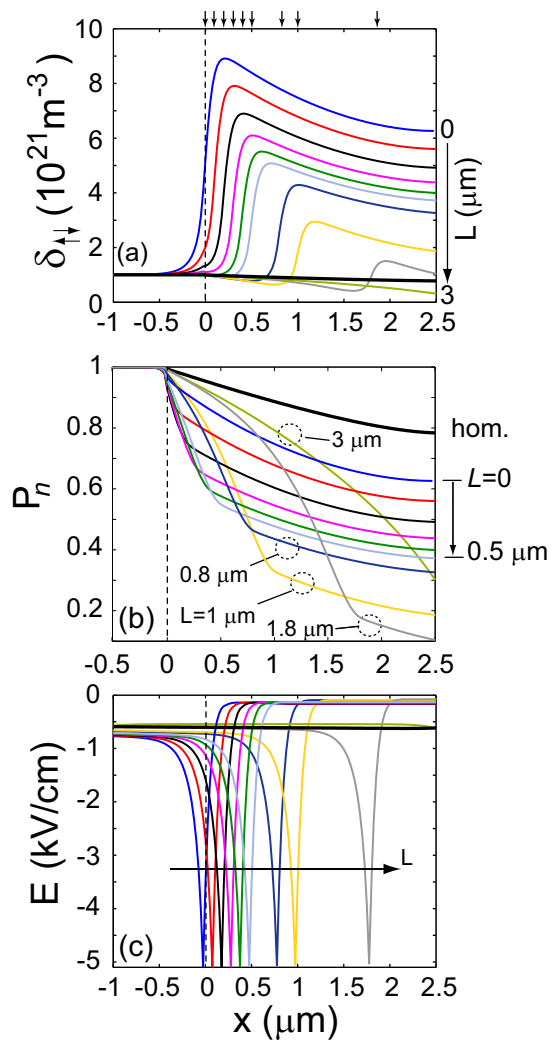}} \\
\caption{(color online) Spin density imbalance $\delta_{\uparrow
\downarrow}$ (a), spin density polarization $P_{n}$ (b), and
electric field distribution (c) for a spin and charge inhomogeneous
structure as depicted in Fig.\ \ref{figure1}, for different channel
lengths, $L=0,0.1,0.2,0.3,0.4,0.5,0.8,1.0,1.8,3.0$ $\mu$m. Curves
are offset such that $x_{p}=0$ in all cases. Parameters are
$N_{L}=10^{21}$ m$^{-3}$, $N_{R}=10^{22}$ m$^{-3}$, $\tau_{{\mathrm
sf}}=1$ ns, $\tau_{{\mathrm m}}=0.1$ ps, $V_{b} = -0.3$ V. Arrows
indicate the position at $x_{d}$. The thick, solid line in the figures corresponds to the charge homogeneous case.} \label{figure3}
\end{figure}
\end{center}

We further clarify the effects of electric field gradients on the
relaxation of nonequilibrium spin distributions. Figure
\ref{figure3} shows the results for $\delta_{\uparrow \downarrow}$
(a), $P_{n}$ (b), and $E(x)$ (c) calculated for $N_{R}=10N_{L}=
10^{22}$ m$^{-3}$ and different channel lengths, $0<L<3$ $\mu$m. We
have offset all curves in space such that $x_{p}$, i.e., the position of the DMS/NMS
interface, coincides. For $L=0$, $\delta _{\uparrow \downarrow}$
shows a sharp increase around the doping interface, followed by a
monotonic, exponential decrease, whereas the spin density
polarization, $P_{n}$, shows an exponential decrease only, with
decay length similar to $\delta_{\uparrow \downarrow}$, but with
values below the corresponding values for the homogeneous system
(thick solid line). This is in agreement with our previous
discussion and agrees with the fact that the electric field [see
Fig.\ \ref{figure3}(c)] in the right-hand side region is
significantly lower than for the homogeneous case [thick solid line
in Fig.\ \ref{figure3}(c)].

For increasing channel lengths the spin imbalance decreases
significantly. In particular, $\delta_{\uparrow \downarrow}$ for the
inhomogeneous structure is {\em smaller} than the corresponding
homogeneous one (thick solid line) for $x<x_{d}$ and a dip-like
feature is formed for large $L$ prior to the interface between the
two doping regions, where $\delta_{\uparrow \downarrow}$ is seen to
increase and reach a peak. Interestingly, this also occurs for
$x<x_{d}-L_{DB}$ for the structures with $L\gg L_{DB}$, i.e., in a
region with very small electric field gradients [see Fig.\
\ref{figure3}(c)]. This illustrates the fact that, although the
electric field gradients in this region are very small [virtually
unobservable on the scale shown in Fig.\ \ref{figure3}(c)], they are
large enough to yield significant influence on the spin density
imbalance. In the spin density polarization, $P_{n}$, the
inhomogeneous electric fields manifest as nonexponential spatial
dependence, which further accentuates the importance of electric
field gradients in the propagation of spin-polarized carriers.

We note that spin accumulation at the interface between regions with
different doping concentrations, such as seen in Fig.\
\ref{figure2}(a), has been previously discussed by Pershin and
Privman.\cite{pershinPRL2003} It was argued that injecting
spin-polarized electrons across a low-to-high doping concentration
interface yields a significant enhancement of the spin polarization.
However, the increase is seen only for $\delta_{\uparrow
\downarrow}$, i.e., the spin density \emph{imbalance}, which the
authors considered in their work. This is, however, only due to the
overall total charge density increase as discussed above. A closer look and comparison to the total charge density, as shown in Fig.\ \ref{figure2}(a) showed that $\delta_{\uparrow\downarrow}$ is suppressed at the interface, which becomes even more evident when considering the actual
spin density \emph{polarization}, $P_{n}$. Moreover, we find that
$P_{n}$ can be enhanced for spin-polarized electron injection across
a high-to-low doping concentration interface, due to the positive
electric field gradients [see dashed-dotted lines in Fig.\
\ref{figure2}(c)].

We would also like to point out the large difference between the
spin density imbalance, $\delta_{\uparrow \downarrow}$, and
polarization, $P_{n}$, in inhomogeneous systems. Previously in the
literature, these two quantities have been used interchangeably,
which certainly is true for charge homogeneous systems. For charge
inhomogeneous systems on the other hand, depending on the measuring
setup at hand, the correct quantity has to be studied and discussed for the optimization and assesment of spin injection, transport and detection schemes. 

We again emphasize that the observed characteristics in the polarization, $P_{n}$, are quite unexpected. Naively, one would expect that, e.g., an increase in doping concentration would result in a simple decrease of $P_{n}$ (which is indeed observed), due to the rapid increase in the total charge density, $n$. However, it is not obvious how $P_{n}$ should vary for different doping concentrations, as in the setup discussed in our paper, and what the magnitude and spatial dependence of $P_{n}$ should be. Our numerical calculations, discussion and analysis above, clearly show that the spatial dependence of the spin density polarization is quite complicated in the presence of inhomogeneous electric fields. 

\section{Conclusions}
\label{conclusions}
The results and discussion presented in our paper predicts that, even weakly, inhomogeneous electric fields
can significantly change the spin diffusion length. Hence, purely
electrical effects may have a profound impact on spin injection,
spin transport, and spin detection, three key ingredients for the
realization of semiconductor spintronics. Our findings should be
relevant to current experimental efforts, as well as theoretical
studies in the field, where we have shown that it is crucial to take
into account inhomogeneous electric fields self-consistently, in
particular around doping interfaces as discussed in our paper. 

\acknowledgments
This work was supported by the Indiana 21st Century Research and
Technology Fund. Numerical calculations were performed using the
facilities at the Center for Computational Nanoscience at Ball State
University.

\newpage

\end{document}